\documentstyle[12pt]{article}
\begin{document}

\title{The Unification and Cogeneration of Dark Matter and Baryonic Matter}
\author{{\bf S.M. Barr} \\
Department of Physics and Astronomy and \\
The Bartol Research Institute \\ University of Delaware \\
Newark, Delaware 19716} \maketitle

\begin{abstract}
In grand unified theories with gauge groups larger than $SU(5)$, the
multiplets that contain the known quarks and leptons also contain
fermions that are singlets under the Standard Model gauge group.
Some of these could be the dark matter of the universe. Grand
unified theories can also have accidental $U(1)$ global symmetries
(analogous to $B-L$ in minimal $SU(5)$) that can stabilize dark
matter. These ideas are illustrated in an $SU(6)$ model.
\end{abstract}

\section{Introduction}

It seems a strange coincidence that the cosmic densities of dark
matter and ordinary baryonic matter are of the same order of
magnitude, given that in most theoretical scenarios they are
generated by unrelated mechanisms involving different particles,
forces, and parameters. This coincidence suggests that the dark
matter and baryonic matter may have been ``cogenerated" in the early
universe, i.e. that the dark matter is a product of the same
processes that created the cosmic baryon asymmetry. There is a
rapidly growing literature studying various ways that this might
have happened
\cite{bcf,adm-hdops,adm-sphaleron,adm-baryonic,adm-other}.

The first papers to propose this possibility \cite{bcf} were based
on the idea that primordial asymmetries in baryon and lepton number
($B$, $L$) were partially converted into an asymmetry in some other
global quantum number (call it $X$) by sphaleron processes
\cite{sphaleron} when the temperature of the universe was above the
weak interaction scale $M_W$. Assuming $X$ to be conserved (or
nearly so) at low temperatures, the lightest particles carrying this
quantum number would be stable and could play the role of dark
matter. What would result from such a scenario is ``asymmetric dark
matter" \cite{nussinov}. Many other scenarios for generating
asymmetric dark matter have been proposed
\cite{adm-hdops,adm-sphaleron,adm-baryonic,adm-other}. In some of
these scenarios ordinary matter and dark matter are converted into
each other by perturbative processes involving higher-dimension
operators \cite{adm-hdops}; and in others by sphalerons (or by both
sphalerons and higher-dimension operators) \cite{adm-sphaleron}. In
some scenarios, the dark matter carries baryon number which
compensates for the non-zero baryon asymmetry of ordinary matter
\cite{adm-baryonic}. And many papers propose still other mechanisms
\cite{adm-other}.

What we suggest here is the possibility that not only are dark
matter and ordinary matter ``cogenerated" but that they are
``unified" in the sense of grand unification. The point is that
grand unification naturally supplies several of the ingredients
needed for the generation of asymmetric dark matter. First, grand
unification based on groups larger than $SU(5)$ involves fermion
multiplets that contain non-Standard-Model fermions that could play
the role of dark matter. In particular, in $SU(N)$ with $N > 5$, the
quark-lepton multiplets contain several fields that are singlets
under the Standard Model group $G_{SM} = SU(2)_L \times SU(3)_c
\times U(1)_Y$. In $SU(6)$ or $SU(7)$ models with three families of
quarks and leptons, for example, anomaly-free sets of fermion
multiplets must contain {\it at least} six Standard-Model-singlet
fermions. For larger groups the number grows rapidly.

Second, it is not uncommon in simple unified models for there to be
global quantum numbers that are ``accidentally" (or
``automatically") conserved, just as $B-L$ is accidentally conserved
in the simplest $SU(5)$ model. Such quantum numbers could play the
role of $X$ that stabilizes the dark matter particles. We shall
illustrate these ideas in a simple $SU(6) \times Z_N$ grand unified
theory (GUT).

Another interesting feature of large unified groups is that they can
contain additional non-abelian factors at low energies (besides
those of $G_{SM}$), whose sphalerons could convert baryons and
leptons into dark matter particles; but we shall not explore that
possibility in this paper. The illustrative model that we shall
describe uses perturbative processes to convert matter and dark
matter into each other \cite{adm-hdops}.

\section{An $SU(6)$ Model}

We shall now present the details of the model. Its symmetry group is
$SU(6) \times Z_N$, where $N$ may be any integer greater than 4. The
fermions of each quark-lepton family consist of the anomaly-free set
of $SU(6)$ multiplets ${\bf 15} + \overline{{\bf 6}} +
\overline{{\bf 6}}$ plus three $SU(6)$ singlets. These are shown in
the left columns of Table I. The fundamental indices of $SU(6)$ are
denoted by the capital latin letters, $A$, $B$, $C$, etc, which run
from 1, ..., 6. Here and throughout the paper, we suppress family
indices. For those fermion multiplets that contain the known quarks
and leptons we use capital $\Psi$. The fermion multiplets denoted by
the letters $\psi$ and $\zeta$ and $\eta$ contain only new fields.

\vspace{0.2cm}

\noindent {\large\bf Table I:} The fermions of a family, and the
Higgs fields.

\vspace{0.2cm}

\begin{tabular}{|l|l|l||l|l|l|}
\hline Fermion field & $SU(6)$ & $Z_N$ & Higgs field & $SU(6)$ & $Z_N$ \\
\hline $\Psi^{[AB]}$ & ${\bf 15}$ & 1 & $\phi^{[AB]}$ & ${\bf 15}$ & 1 \\
\hline  $\Psi_A$ & $\overline{{\bf 6}}$ & 1 & $\phi_A$ & $\overline{{\bf 6}}$ & 1 \\
\hline $\Psi$ & ${\bf 1}$ & 1  &  $H_A$ & $\overline{{\bf 6}}$ & $\omega$ \\
\hline $\psi'_A$ &
$\overline{{\bf 6}}$ & $\omega^*$  & & & \\
\hline  $\eta$ & ${\bf 1}$ & $\omega$ &  $\Omega^A_B$ & ${\bf 35}$ & $\omega^*$ \\
\hline $\zeta$ & ${\bf 1}$ & $\omega^2$ & & & \\
\hline
\end{tabular}

\vspace{0.5cm }

\noindent The right columns of Table I show the Higgs fields and
their $SU(6) \times Z_N$ transformation properties. The known fields
all transform trivially under $Z_N$. The gauge symmetry breaking
occurs in three stages:

(1) At the unification scale $M_{GUT}$, $SU(6)$ is broken to
$[SU(3)_c \times SU(2)_L \times U(1)_Y] \times U(1)_6$ (which is
contained in the $SU(5) \times U(1)_6$ subgroup of $SU(6)$). This
breaking is done by an adjoint Higgs field $\Omega^A_B$, whose
vacuum expectation value (VEV) points in a direction that is a
linear combination of the weak hypercharge generator $Y/2 = diag
(\frac{1}{2}, \frac{1}{2}, - \frac{1}{3}, -\frac{1}{3},
-\frac{1}{3}, 0)$ and the $U(1)_6$ generator $T_6 = diag
(-\frac{1}{5}, -\frac{1}{5}, -\frac{1}{5}, -\frac{1}{5},
-\frac{1}{5}, 1)$. We shall always denote an $SU(5)$ index (which
takes values 1,2,3,4,5) by $\alpha$, $\beta$, etc.; an $SU(2)_L$
index (which takes values 1,2) by $i$, $j$, etc.; and an $SU(3)_c$
index (which takes values 3,4,5) by $a$, $b$, etc. Thus, for
example, $\psi_A = (\psi_{\alpha}, \psi_6) = (\psi_i, \psi_a,
\psi_6)$.

(2) The $U(1)_6$ is broken at a scale $M'$, which is somewhat larger
than a TeV, by the fundamental Higgs multiplet $H_A$, whose VEV
points in the 6 direction, i.e. $\langle H_6 \rangle \sim M'$. We
assume that the mass of $H_6$ is of order $M'$, but that its other
components all have mass of order $M_{GUT}$. (This is the usual kind
of split-multiplet problem of GUTs, analogous to the well-known
``doublet-triplet splitting problem".) Below $M'$, the gauge group
is just the Standard Model group $G_{SM} = SU(3)_c \times SU(2)_L
\times U(1)_Y$. There is, of course, a $Z'$ gauge boson whose mass
is of order $M'$ that couples to $T_6$.

(3) The electroweak breaking is done by the two Higgs multiplets
denoted by $\phi$, which contain the $SU(2)_L$ doublets $\phi_i$ and
$\phi^*_{{i6}}$. We assume that one linear combination of these two
doublets is tuned to be light (i.e. of order the weak scale in mass)
and is the Standard Model Higgs doublet $\Phi_i$ that obtains a VEV,
while all the other components of $\phi_A$ and $\phi^{[AB]}$ have
superheavy mass. (In particular, the colored components, which can
mediate proton decay, have mass of order $M_{GUT}$.)

Under the subgroup $SU(5) \subset SU(6)$, the non-singlet fermions
of Table I decompose as follows

\begin{equation}
\begin{array}{llll}
\Psi^{[AB]} & \rightarrow \Psi^{[\alpha \beta]} & + & \Psi^{\alpha
6}
\\
{\bf 15} & \rightarrow {\bf 10} & + & {\bf 5}
\\
& & & \\
\Psi_A & \rightarrow \Psi_{\alpha} & + & \Psi_6  \\
\overline{{\bf 6}} & \rightarrow \overline{{\bf 5}} & +
& {\bf 1} \\ & & & \\
\psi'_A & \rightarrow \psi'_{\alpha} & + & \psi'_6 \\
\overline{{\bf 6}} & \rightarrow \overline{{\bf 5}} & + & {\bf 1}
\\ & & \end{array}
\end{equation}

\noindent The Standard Model quarks and leptons are the
$\Psi^{[\alpha \beta]} = {\bf 10}$ and $\Psi_{\alpha} =
\overline{{\bf 5}}$. The extra $\overline{{\bf 5}} + {\bf 5}$ pair
of $SU(5)$ will ``mate" to get $O(M')$ masses. Specifically, the
${\bf 5} = \Psi^{[\alpha 6]}$ will obtain mass with $\overline{{\bf
5}} = \psi'_{\alpha}$ through a term $(\Psi^{[\alpha 6]}
\psi'_{\alpha}) \langle H_6 \rangle$, as will be seen. There are
also two $SU(5)$-singlet (and thus $G_{SM}$-singlet) fermions in the
multiplets shown in Eq. (1). In order for these to get mass, we
introduce gauge-singlets fermions denoted $\eta$ and $\zeta$ that
will mate with them to get Dirac masses.

The most general renormalizable Yukawa couplings allowed by $SU(6)
\times Z_N$ are of the following forms (we suppress family indices):

\begin{equation}
\begin{array}{rl}
{\cal L}_{Yukawa} & = \; {\cal L}_{SM} + {\cal L}_{F \overline{F}} + {\cal L}_{singlet} \\ & \\
{\cal L}_{SM} & = \; Y_u (\Psi^{[AB} \Psi^{CD}) \phi^{EF]}
 + Y_d (\Psi^{[AB]} \Psi_A) \phi_B + \; Y_{\nu} (\Psi_A \Psi) \phi^{*A} + M_R (\Psi \Psi) \\ & \\
{\cal L}_{F \overline{F}} & = \; f_1 (\Psi^{[AB]} \psi'_A) H_B
\\ & \\
{\cal L}_{singlet} & = f_2 (\Psi_A \eta) H^{*A} + \; f_3 (\psi'_A
\zeta) H^{*A} + f_4 (\psi'_A \eta) \phi^{*A}.
\\ \end{array}
\end{equation}

\noindent The first three terms in ${\cal L}_{SM}$ have the effect
of coupling the Standard Model Higgs doublet (which is a mixture of
$\phi_i$ and $\phi^*_{i6}$) to the known quarks and leptons (which,
as noted above, are contained in the multiplets that are denoted by
capital $\Psi$'s). The term ${\cal L}_{F \overline{F}} = f_1
(\Psi^{[AB]} \psi'_A) H_B$ gives $O(M')$ mass to the ``extra" ${\bf
5} + \overline{{\bf 5}}$ pair of fermions, as already mentioned. The
terms in ${\cal L}_{singlet}$ couple the Standard-Model-singlet
fermions $\Psi_6$ and $\psi'_6$ (see Eq. (1)) to the gauge-singlet
fermions denoted $\eta$ and $\zeta$ so that all the singlet fermions
can get mass. All these Yukawa terms and the masses that come from
them will be examined in more detail shortly.

The most general Yukawa terms allowed by $SU(6) \times Z_N$ (shown
in Eq. (2)) and the most general Higgs potential allowed by $SU(6)
\times Z_N$ (which, incidentally, includes terms such as $\phi^{AB}
\phi_A H_C \Omega^C_B$) happen ``accidentally" to be invariant under
a global $U(1)$ symmetry, whose generator we will call $T$. The $T$
charges of the various multiplets in the model are as follows (given
in parentheses): $\Psi^{[AB]} (1)$, $\Psi_A (-\frac{1}{2})$, $\Psi
(0)$, $\psi'_A (-\frac{7}{2})$, $\eta (3)$, $\zeta (6)$,
$\phi^{[AB]} (-2)$, $\phi_A (-\frac{1}{2})$, $H_A (\frac{5}{2})$,
and $\Omega^A_B (0)$. This $U(1)_T$ symmetry is unbroken by
GUT-scale VEVs, but is spontaneously broken at the scale $M'$ by
$\langle H_6 \rangle$, which also breaks the gauge group $U(1)_6$,
leaving an unbroken global symmetry $U(1)_X$, whose generator is
given by

\begin{equation}
\begin{array}{l}
\\ X = \frac{1}{3} T + \frac{5}{6} T_6.
\\ \\ \end{array}
\end{equation}

\noindent This generator $X$ will play a crucial role in what
follows as the quantum number that stabilizes dark matter. It is
analogous to the conserved global symmetry $B-L$ in minimal $SU(5)$
models, where $B-L$ is a linear combination of a global charge that
is accidentally conserved by the Yukawa couplings and a gauge
generator (specifically, the weak hypercharge $Y$). Here, both $X$
and $B-L$ are conserved by low-energy couplings and VEVs.

All the Higgs fields in Table I have $O(M_{GUT})$ masses except (a)
the Standard Model doublet $\Phi_i$, which is a linear combination
of the doublets $\phi_i$ and $ \phi^*_{[i6]}$, and has mass of order
100 GeV, and (b) the Standard-Model-singlet Higgs $H_6$, which has
mass of order $M' > 1$ TeV. We will call $\Phi_i$ and $H_6$ ``light
Higgs fields" to distinguish them from ``superheavy Higgs fields".
It is easy to check explicitly that the light Higgs fields $\Phi_i$
and $H_6$ are neutral under $X$ (as must be the case, of course, if
$X$ is left unbroken by their VEVs). Therefore, when fermions absorb
or emit these light Higgs fields they do not change their $X$
charge. The same is true of the emission and absorption of light
gauge bosons (i.e. those of $SU(3)_c \times SU(2)_L \times U(1)_Y
\times U(1)_6$).

In other words, except through extremely slow processes mediated by
bosons with superheavy masses, fermions do not change their values
of $X$. It is therefore very useful to classify the fermions of this
model by their $X$ charges. This is done in Table II.

Note that Table II also lists quantum numbers called $B_0$, $L_0$,
$B_1$, $L_1$, $L_2$. These are defined as follows: $B_n \equiv B
\delta_{|X|n}, L_n \equiv L \delta_{|X| n}$. In other words we
define separate baryon and lepton numbers for each $|X|$ sector. For
example, a lepton with $X=-2$ has $L_2 = 1$, but $L_0 = L_1 = 0$.
(It should be noted that $B = B_0 + B_1$, $L= L_0 + L_1 + L_2$, and
$X = 3B_1 - L_1 - 2L_2$.) The reason for defining these baryon and
lepton numbers is that the emission and absorption of ``light Higgs
fields" and ``light gauge bosons" do not change the values of $B$,
$L$, and $X$ of a fermion and therefore also separately conserve the
quantum numbers $B_n$ and $L_n$. This will be important in our later
analysis. Later we shall introduce four-fermion operators, generated
by the exchange of very heavy bosons, that conserve $B$, $L$ and $X$
but violate $B_n$ and $L_n$. Such processes will be needed to
re-distribute particle asymmetries, i.e. convert a primordial
asymmetry in one global charge into the other global charges, so
that matter and dark matter asymmetries will end up being related to
each other.

\newpage

\noindent {\large\bf Table II:} The fermions and Higgs fields,
classified by their $X$ values (0, $\pm 1$, $\pm 2$), where $X =
\frac{1}{3} T + \frac{5}{6} T_6$. The baryon and lepton numbers are
defined by $B_n \equiv B \delta_{|X|n}, L_n \equiv L \delta_{|X|
n}$. \vspace{0.2cm}

\begin{tabular}{|ll||r|r|r||r|r|r|r|r|}
\hline
{\bf Field} & & $X$ & $T$ & $T_6$ & $B_0$ & $L_0$ & $B_1$ & $L_1$ & $L_2$ \\
\hline \hline
$\Psi^{[\alpha \beta]}$ & $\rightarrow \Psi^{[ab]} = u^c$ & 0 & $1$ & $-\frac{2}{5}$ & $-\frac{1}{3}$ & 0 & 0 & 0 & 0 \\
& $\rightarrow \Psi^{[ai]} = Q$ & 0 & $1$ & $-\frac{2}{5}$ & $\frac{1}{3}$ & 0 & 0 & 0 & 0  \\
& $\rightarrow \Psi^{[12]} = \ell^c$ & 0 & $1$ & $-\frac{2}{5}$ & 0 & $-1$ & 0 & 0 & 0 \\
\hline $\Psi_{\alpha}$ & $\rightarrow \Psi_a = d^c$ & 0 & $-\frac{1}{2}$ &
$\frac{1}{5}$ & $-\frac{1}{3}$ & 0 & 0 & 0 & 0 \\
& $\rightarrow \Psi_i = L$ & 0 & $-\frac{1}{2}$ & $\frac{1}{5}$ & 0 & $1$ & 0 & 0 & 0 \\
\hline $\Psi$ & $\rightarrow \Psi = N^c$ & 0 & 0 & 0 & 0 & $-1$ & 0 & 0 & 0 \\
\hline \hline $\Psi^{[\alpha 6]}$ & $\rightarrow \Psi^{[a6]}$
& $1$ & $1$ & $\frac{4}{5}$ & 0 & 0 & $\frac{1}{3}$ & 0 & 0 \\
& $\rightarrow \Psi^{[i6]}$ & $1$ & $1$ & $\frac{4}{5}$ & 0 & 0 & 0 & $-1$ & 0 \\
\hline $\psi'_{\alpha}$ & $\rightarrow \psi'_a$ &
$-1$ & $-\frac{7}{2}$ & $\frac{1}{5}$ & 0 & 0 & $-\frac{1}{3}$ & 0 & 0 \\
& $\rightarrow \psi'_i$ & $-1$ & $-\frac{7}{2}$ & $\frac{1}{5}$ & 0 & 0 & 0 & $1$ & 0 \\
\hline $\Psi_6$ & & $-1$ & $-\frac{1}{2}$ & $-1$ & 0 & 0  & 0 & $1$ & 0 \\
\hline $\eta$ &  & $1$ & $3$ & 0 & 0 & 0 & 0 & $-1$ & 0 \\
\hline \hline $\psi'_6$ & & $-2$ & $-\frac{7}{2}$ & $-1$ & 0 & 0 & 0 & 0 & $1$ \\
\hline $\zeta$ & & $2$ & $6$ & 0 & 0 & 0 & 0 & 0 & $-1$ \\
\hline \hline $\phi_{\alpha}$ & $\rightarrow \phi_i$ & 0 & $-\frac{1}{2}$ & $\frac{1}{5}$ & 0 & 0 & 0 & 0 & 0 \\
\hline $\phi^{[\alpha 6]}$ & $\rightarrow \phi^{[i6]}$ & 0 & $-2$ & $\frac{4}{5}$ & 0 & 0 & 0 & 0 & 0 \\
\hline $H_6$ & & 0 & $\frac{5}{2}$ & $-1$ & 0 & 0 & 0 & 0 & 0 \\
\hline
\end{tabular}

\vspace{0.5cm}

Observe that the fermions with $X=0$ are just the known quarks and
leptons of the Standard model. Since the light Higgs fields have
$X=0$, their Yukawa couplings only couple these Standard Model
fermions to each other and give them Dirac masses with each other.
These couplings come from the terms ${\cal L}_{SM}$ in Eq. (2).
Specifically, the $Y_u$ term gives mass to up-type quarks via $Y_u
(\Psi^{[ab]} \Psi^{[c1]}) \langle \phi^{[26]} \rangle \propto Y_u
(u^c u) v$. The $Y_d$ term gives mass to down-type quarks and
charged leptons via $Y_d (\Psi^{[a2]} \Psi_a + \Psi^{[21]} \Psi_1)
\langle \phi_2 \rangle \propto Y_d (d d^c + \ell^+ \ell^-) v$. The
$Y_{\nu}$ term gives the Dirac neutrino masses via $Y_{\nu} (\Psi_2
\Psi) \langle \phi^{*2} \rangle \propto Y_{\nu} (\nu N^c) v$. And
the $M_R$ term gives the superlarge Majorana masses to the
right-handed neutrinos: $M_R (\Psi \Psi) = M_R (N^c N^c)$.

As can be seen from Table II, the sector of fermions with $X= \pm 1$
contains (for each family) a ${\bf 5}$ and $\overline{{\bf 5}}$'s of
$SU(5)$, namely $\Psi^{[\alpha 6]}$ and $\psi'_{\alpha}$. These
``mate" to obtain masses of $O(M')$ via the Yukawa term in ${\cal
L}_{F \overline{F}}$, which gives $f_1 (\Psi^{[\alpha 6]}
\psi'_{\alpha}) \langle H_6 \rangle$.

The $X= \pm 1$ sector also contains (for each family) the singlet
fermions $\Psi_6$ with $X = -1$, and $\eta$ with $X=1$. The term
$f_2 (\Psi_6 \eta) H^{*6}$ in ${\cal L}_{singlet}$ couples them
together into massive Dirac particles. (Note that the gauge-singlet
fermions $\eta$ have been introduced into the model just to give
mass to $\Psi_6$, which otherwise would remain massless.) The term
$f_4 (\psi'_6 \eta) \phi^{*6}$ in ${\cal L}_{singlet}$ has the
effect of mixing these singlet fermions with neutrinos in the ${\bf
5} + \overline{{\bf 5}}$, so that the neutral fermions in the $X =
\pm 1$ sector actually have a $2 \times 2$ mass matrix (actually $6
\times 6$ if one takes into account that there are three families)
of the following form:

\begin{equation}
\left( \psi'_2, \Psi_6 \right) \; \left(
\begin{array}{cc} f_1 \langle H_6 \rangle & f_4 \langle \phi^{*2} \rangle  \\
0 & f_2 \langle H^{*6} \rangle \end{array} \right) \; \left(
\begin{array}{c} \Psi^{[26]} \\ \eta \end{array} \right).
\end{equation}

\noindent If all the Yukawa couplings in Eq. (4) were of order 1,
then all the masses of the fermions in the $X = \pm 1$ sector would
be of $O(M')$. There have to be, however, some particles to play the
role of dark matter. Since in this scenario the present number
densities of dark matter particles and baryons will turn out to be
of the same order of magnitude, the mass $m_{DM}$ of the dark matter
particles should be roughly of order 1 GeV. There are various ways
this can be the case. One simple way is that the Yukawa couplings
that we have denoted $f_2$ in Eqs. (2) and (4) are of order $(1 \;
{\rm GeV})/M' < 10^{-3}$. We shall assume this to be true and also
assume that the Yukawa couplings denoted $f_1$ and $f_3$ are
significantly larger than $f_2$. In that case, the lightest $X \neq
0$ fermions are the Dirac fermions made up of the gauge singlets
$\Psi_6$ and $\eta$. These will be the dark matter particles of the
model. We will denote these dark matter particles sometimes as
$(\Psi_6, \eta)$. (As noted, and as can be seen from Eq. (4), these
mix with angle $O(v/M')$ with weak-doublet neutrinos of mass
$O(M')$. Thus the dark matter particles have $O \left( \left(
\frac{v}{M'} \right)^2 \frac{f_2}{f_1} \right)$ couplings to the
Standard Model $Z$ boson.)

We come, finally, to the $X= \pm 2$ sector of fermions. This
consists (for each family) of a Standard-Model-singlet fermions with
$X=-2$ (namely, $\psi'_6$), and with $X=+2$ (namely $\zeta$). The
term $f_3 (\psi'_A \zeta) H^{*A}$ in ${\cal L}_{singlet}$ couples
these together to make massive Dirac particles. (Note that the
gauge-singlet fermions denoted by the letter $\zeta$ have been
introduced into the model just to give mass to the $X=-2$ fermions.)

\section{The Processes that Redistribute Asymmetries}

As noted before, there must be processes that conserve $X$ and $B-L$
but violate $B_n$ and $L_n$ in order to redistribute asymmetries in
quantum numbers and thus relate the matter and dark matter
asymmetries. Such processes are needed for other reasons as well.
For example, they are needed to allow the colored $X\neq 0$
particles (i.e. those with $B_1 \neq 0$) to decay. (Such particles
if stable and light would have been seen at accelerators, and if
stable and heavy would contribute too much to the dark matter
density of the universe.) These $\Delta B_1 \neq 0$ decays can be
relatively slow, as long as they occur early enough not to interfere
with primordial nucleosynthesis.

The processes that we postulate to violate $B_n$ and $L_n$ are very
simple. They are given by four-fermion operators of the form
$(\overline{\psi'}^A \psi'_B) (\overline{\Psi}^B \Psi_A)$, and more
precisely by

\begin{equation}
\begin{array}{ll}
{\cal O}_1 = (M_1)^{-2} (\overline{\psi'}^6 \psi'_i)
(\overline{\Psi}^i \Psi_6), \;\; & \Delta (B_0, L_0, B_1, L_1, L_2) = (0, -1, 0, 2, -1),  \\
& \\
{\cal O}_2 = (M_2)^{-2} (\overline{\psi'}^6 \psi'_a)
(\overline{\Psi}^a \Psi_6), \;\; & \Delta (B_0, L_0, B_1, L_1, L_2) = (\frac{1}{3}, 0, -\frac{1}{3} , 1, -1), \\
& \\
{\cal O}_3 = (M_3)^{-2} (\overline{\psi'}^a \psi'_i)
(\overline{\Psi}^i \Psi_a), \;\; & \Delta (B_0, L_0, B_1, L_1, L_2)
= (-\frac{1}{3}, -1, \frac{1}{3}, 1, 0). \\ \end{array}
\end{equation}

\noindent It is easy to check that these conserve $B = B_0 + B_1$,
$L= L_0 + L_1 + L_2$ and $X=3B_1 - L_1 - 2L_2$. These operators can
arise in a simple way from integrating out heavy fields as follows.
Suppose there is a boson $\tilde{\phi}_A$ and gauge-singlet fermions
$\tilde{\Psi}$ and $\tilde{\eta}$. Suppose $\tilde{\phi}_A$,
$\tilde{\Psi}$, and $\tilde{\eta}$ have the same $SU(6) \times Z_N$,
quantum numbers as the particles $\phi_A$, $\psi$ and $\eta$,
respectively. We assume that $\tilde{\phi}_\alpha$ and
$\tilde{\eta}$ have mass of a scale we will call $M_{\Delta}$, where
$M_{\Delta} \gg M'$, and that $\tilde{\phi}_6$ has mass of order 1
to 10 GeV, while $\tilde{\Psi}$ is massless. (It is also assumed
that $\tilde{\phi}_A$ has vanishing VEV.) With these quantum
numbers, these particles have the Yukawa couplings

\begin{equation}
{\cal L}_{\Delta} = f (\Psi_A \tilde{\Psi}) \tilde{\phi}^{*A} + f'
(\psi'_A \tilde{\eta}) \tilde{\phi}^{*A}.
\end{equation}

\noindent Note the similarity to the terms in Eq. (2) with
coefficients $Y_{\nu}$ and $f_4$. We can ensure that these are the
{\it only} Yukawa couplings that $\tilde{\phi}_A$, $\tilde{\Psi}$,
and $\tilde{\eta}$ possess, either by assigning them suitable $Z_N$
charges, or, even more simply, by positing another $Z_M$ symmetry
under which $\tilde{\phi}_A \rightarrow z \tilde{\phi}_A$,
$\tilde{\Psi} \rightarrow z \tilde{\Psi}$, and $\tilde{\eta}
\rightarrow z \tilde{\eta}$, while all other fields transform
trivially. (We can assign the massless fermion $\tilde{\Psi}$ lepton
numbers $L_0 = 0$, $L_1 = 1$, $L_2 =0$, with the light boson
$\tilde{\phi}_6$ having all baryon and lepton numbers zero.)

Then the box diagram shown in Fig. 1 gives rise to the operators in
Eq. (5), with $M_1, M_2, M_3 \sim M_{\Delta}$. Note that the
operator ${\cal O}_3$ directly gives the decay $\psi'_a
\longrightarrow \Psi_a + \overline{\Psi}^2 + \psi'_2$. The initial
particle is an $X =-1$ antiquark with mass of $O(M')$. The first two
final state particles are ordinary Standard Model particles, namely
an antiquark ($d^c_L$) and a lepton ($\nu_L$). The third final state
particle is $\psi'_2$, which mixes with the dark matter particles
$(\Psi_6, \eta)$ as can be seen from Eq. (4). In fact, the operators
in Eq. (5) allow all fermions of the model with masses of order $M'$
to decay ultimately to ordinary quarks and leptons and the dark
matter particles $(\Psi_6,\eta)$.

\vspace{0.2cm}

\begin{picture}(360,180)
\thicklines \put(90,60){\line(1,0){25}}
\put(150,60){\vector(-1,0){35}} \put(150,60){\vector(1,0){35}}
\put(185,60){\line(1,0){25}} \put(210,60){\line(1,0){25}}
\put(270,60){\vector(-1,0){35}} \put(90,120){\vector(1,0){35}}
\put(125,120){\line(1,0){25}} \put(150,120){\line(1,0){25}}
\put(210,120){\vector(-1,0){35}} \put(210,120){\vector(1,0){35}}
\put(245,120){\line(1,0){25}} \put(150,120){\vector(0,-1){35}}
\put(150,60){\line(0,1){25}} \put(210,60){\vector(0,1){35}}
\put(210,95){\line(0,1){25}} \put(110,47){$\psi'_A$}
\put(175,47){$\tilde{\eta}$} \put(230,47){$\psi'_B$}
\put(155,85){$\tilde{\phi}_A$} \put(215,85){$\tilde{\phi}_B$}
\put(110,127){$\Psi_A$} \put(175,127){$\tilde{\Psi}$}
\put(230,127){$\Psi_B$} \put(160,20){{\bf Fig. 1}}
\end{picture}

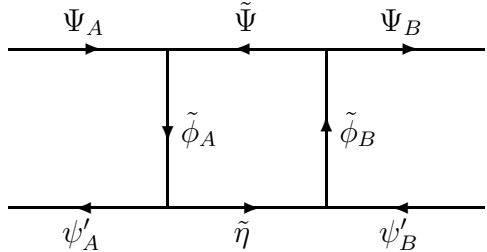
\noindent {\bf Figure 1:} The diagram that gives the operators
${\cal O}_1$ (if $A=6$ and $B=i$), ${\cal O}_2$ (if $A=6$ and
$B=a$), and ${\cal O}_3$ (if $A=a$ and $B=i$).

\section{The Cosmological Scenario}

Now let us outline the sequence of events in the early universe that
generate the current baryon and dark matter abundances in this
model.

Stage 1, which happens when the universe is above a superlarge
temperature $T_{LG}$, is the genesis of an asymmetry in lepton
number. Often this is assumed to happen through the decays of
superheavy right-handed neutrinos (here called $\Psi$)
\cite{rhn-leptogenesis}. However, as will be seen, this will not
lead to the generation of any dark matter asymmetry in the present
scenario, since the right-handed neutrinos have $X=0$. We will
therefore assume that the primordial asymmetry is of fermions that
carry both $L$ and $X$. In particular, we shall look at the case in
which an asymmetry of the fermions $\zeta$ is generated. How this
might happen will be discussed later.

Stage 2 is the period $T_{\Delta} < T < M_{\Delta}$, during which
both $SU(2)_L$ sphaleron processes and the scattering processes
involving the operators ${\cal O}_1$, ${\cal O}_2$ and ${\cal O}_3$
are in equilibrium. The freeze-out temperature $T_{\Delta}$ of those
scattering processes is given roughly by $T_{\Delta} \sim M_{\Delta}
\left( \frac{16 \pi g^{1/2} M_{\Delta}}{M_{P \ell}} \right)^{1/3}$,
where $g$ is the effective number of massless particle species at
$T_{\Delta}$. We shall assume that $M_{\Delta}$ is large enough that
$T_{\Delta} > M'$, the scale at which $U(1)_6$ breaks and the
non-Standard-Model quarks and leptons get their mass. During stage
2, the initial asymmetry in $\zeta$ (and thus in $L$ and $X$) is
reshuffled by sphalerons and scattering processes among the various
particle types, leading to asymmetries in all of the quantum numbers
$B_0$, $L_0$, $B_1$, $L_1$, and $L_2$ that are of the same order.
These will be computed shortly.

Stage 3 is the period $T_{dec} < T < T_{\Delta}$, where $T_{dec}$ is
the temperature at which the particles with mass of $O(M')$ decay.
When the temperature falls below $T_{\Delta}$, the relative values
of $B_0$, $L_0$, $B_1$, $L_1$, and $L_2$ freeze. Sphaleron processes
continue until some temperature $T_{sph} \sim 200$ GeV, but do not
affect these ratios, which change again only when the fermions with
mass of order $O(M')$ decay (out of equilibrium) via the operators
${\cal O}_1$, ${\cal O}_2$, and ${\cal O}_3$ at $T_{dec} \sim M'
\left( \frac{M^{\prime 3}}{96 \pi^2 T_{\Delta}^3} \right)^{1/2}$. It
should be noted that when $T$ falls below $M'$ (which happens during
stage 3, since $T_{dec} < M' < T_{\Delta}$), annihilations (mediated
by $SU(3)_c \times SU(2)_L \times U(1)_Y \times U(1)_6$ gauge
interactions wipe out almost all the particles of $O(M')$ except for
the asymmetric components. (For example, if the asymmetry in $B_1$
is positive, the density of $B_1 <0$ particles will be driven to a
value much less than that of $B_1
>0$ particles.)

Stage 4 is the period when $T < T_{dec}$. At this point, the
particles with mass of $O(M')$ decay out of equilibrium via the
operators ${\cal O}_1$, ${\cal O}_2$, and ${\cal O}_3$. These decays
violate $B_0$, $L_0$, $B_1$, $L_1$, and $L_2$, and so again
reshuffle the ratios of these quantum numbers. In particular, they
set $B_1$ and $L_2$ to zero; but they leave a non-zero $L_1$ in the
form of the dark matter particles $(\Psi_6, \eta)$, which we will
compute below. At this point, the only remaining fermions are the
known Standard Model quarks and leptons, the dark matter particles
$(\Psi_6, \eta)$, and the massless $\tilde{\Psi}$ particles (whose
contribution to the radiation density at the time of nucleosynthesis
is equivalent to less than half of a neutrino species).

We assume that $T_{dec} > m_{DM} \sim 1$ GeV. (This means, for
example, that if $M' \sim 1$ TeV, then $T_{\Delta}$ must be also be
of order 1 TeV, while if $M' \sim 10$ TeV, then $T_{\Delta}$ must be
less than about 100 TeV.) When $T$ falls below $m_{DM}$ the dark
matter particles $(\Psi_6, \eta)$ start to annihilate with their
antiparticles through the diagrams shown in Fig. 2. Since the
$\tilde{\phi}_6$ have been assumed to be light (of order 1 to 10
GeV) these annihilations efficiently reduce the number of dark
matter anti-particles to much below the number of dark matter
particles. That is, the dark matter is asymmetric.

\vspace{0.2cm}

\begin{picture}(360,180)
\thicklines \put(120,60){\line(1,0){25}}
\put(180,60){\vector(-1,0){35}} \put(180,60){\vector(1,0){35}}
\put(215,60){\line(1,0){25}} \put(120,120){\vector(1,0){35}}
\put(155,120){\line(1,0){25}} \put(180,120){\line(1,0){25}}
\put(240,120){\vector(-1,0){35}} \put(180,120){\vector(0,-1){35}}
\put(180,60){\line(0,1){25}} \put(140,47){$\Psi_6$}
\put(205,43){$\tilde{\Psi}$} \put(185,85){$\tilde{\phi}_6$}
\put(140,127){$\Psi_6$} \put(205,127){$\tilde{\Psi}$}
\put(160,20){{\bf Fig. 2}}
\end{picture}

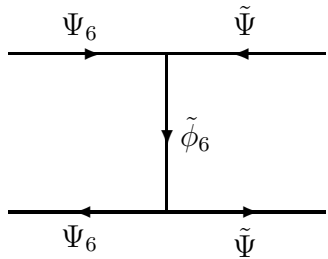
\noindent {\bf Figure 2:} The diagram by which dark matter particles
and antiparticles can annihilate into massless fermion antifermion
pairs: $\Psi_6 + \overline{\Psi}^6 \rightarrow \tilde{\Psi} +
\overline{\tilde{\Psi}}$.

\section{Computing the Relative Asymmetries}

We now calculate the matter and dark matter asymmetries that result
in this model. During stage 2, all of the fermions of the model
except $\Psi$ and $\tilde{\eta}$ are relativistic, so we can neglect
their masses. Call the fermion chemical potentials $\mu_{q_0}$,
$\mu_{\ell_0}$, $\mu_{q_1}$, $\mu_{\ell_1}$, and $\mu_{\ell_2}$, for
the quarks and leptons having various values of $X$. From the fact
that the sphaleron processes and the scattering processes that
involve the operators ${\cal O}_1$, and ${\cal O}_3$ are in
equilibrium, one has the following relations

\begin{equation}
\begin{array}{ll}
{\rm sphaleron:} & 0 = 3 \mu_{q_0} + \mu_{\ell_0}, \\ & \\
{\cal O}_1: & 0 = \mu_{\ell_0} - 2 \mu_{\ell_1} + \mu_{\ell_2}, \\ & \\
{\cal O}_3: & 0 = \mu_{q_0} + \mu_{\ell_0} - \mu_{q_1} -
\mu_{\ell_1}.
\end{array}
\end{equation}

\noindent The equilibrium of the scattering processes involving
${\cal O}_2$ does not give an independent relation. From Table I, it
can be seen that the number of species of fermions of each type
(i.e. the statistical weight), counting the number of families,
colors and polarizations, is given by $g(q_0) = 3 \cdot 3 \cdot 4 =
36$, $g(\ell_0) = 3 \cdot 1 \cdot 3 = 9$, $g(q_1) = 3 \cdot 3 \cdot
2 = 18$, $g(\ell_1) = 3 \cdot 1 \cdot 6 + 1 = 19$, and $g(\ell_2) =
3 \cdot 1 \cdot 2 = 6$. (The extra 1 appearing in $g(\ell_1)$ is due
to the fermion $\tilde{\Psi}$.) In a comoving volume of the
universe, the asymmetries in the quantum numbers are related to the
chemical potentials by the relation $N_i \propto g_i \mu_i T^2
R(T)^3$, where $R(T)$ is the scale factor of the universe when the
temperature is $T$. Therefore, one has

\begin{equation}
\begin{array}{rl}
B_0 = \frac{1}{3} Q_0 & = 12 \mu_{q_0} K, \\ & \\
L_0 & = 9 \mu_{\ell_0} K, \\ & \\
B_1 = \frac{1}{3} Q_1 & = 6 \mu_{q_1} K, \\ & \\
L_1 & = 19 \mu_{\ell_1} K, \\ & \\
L_2 & = 6 \mu_{\ell_2} K, \end{array}
\end{equation}

\noindent where $K$ depends on the temperature and volume.
Therefore, from Eqs. (7) and (8), one has that during stage 2

\begin{equation}
\begin{array}{l}
0 = \frac{1}{4} B_0 + \frac{1}{9} L_0, \\ \\
0 = \frac{1}{12} B_0 + \frac{1}{9} L_0 - \frac{1}{6} B_1 - \frac{1}{19} L_1, \\ \\
0 = \frac{1}{9} L_0 - \frac{2}{19} L_1 + \frac{1}{6} L_2.
\end{array}
\end{equation}

\noindent We assume that during stage 1 primordial asymmetries were
generated in the quantum numbers $X$ and $B-L$. After stage 1,
however, these quantum numbers are conserved. Thus we have two
further relations

\begin{equation}
\begin{array}{rlr}
X & = 3 B_1 - L_1 - 2 L_2 & = a, \\ & & \\
B-L & = B_0 + B_1 - L_0 - L_1 - L_2 & = b,
\end{array}
\end{equation}

\noindent where $a$ and $b$ are constants. Eqs. (9) and (10) can be
solved to obtain,

\begin{equation}
\begin{array}{l}
B_0 = - \frac{4}{9 \cdot 119} \left( 37 a - 61 b \right),  \\ \\
L_0 = \frac{1}{119} \left( 37 a - 61 b \right), \\ \\
B_1 = \frac{4}{9 \cdot 119} \left( 83 a - 50 b \right), \\ \\
L_1 = - \frac{19}{3 \cdot 119} \left( a + 8 b \right), \\ \\
L_2 = \frac{1}{3 \cdot 119} \left( -86 a + 26 b \right).
\end{array}
\end{equation}

In stage 4, after $T$ has fallen below $T_{dec}$, the particles with
mass of order $M'$ decay into the ordinary quarks and leptons of the
Standard Model, plus the dark matter fields $\Psi_6$. We will assume
for ease of discussion that the heaviest $O(M')$ particles are those
with $B_1 \neq 0$ followed by those with $L_2 \neq 0$. Then the $B_1
\neq 0$ particles will decay via the operator ${\cal O}_3$. From Eq.
(5) one sees that these decays will change the particle asymmetries
in the proportions $\Delta B_0 = - \Delta B_1$, $\Delta L_0 = -3
\Delta B_1$, and $\Delta L_1 = 3 \Delta B_1$. Therefore, if $B_1
\longrightarrow B'_1 = B_1 + \Delta B_1 = 0$, one has

\begin{equation}
\begin{array}{l}
B'_0 = B_0 + \Delta B_0 = \frac{2}{119} \left( a + 8 b \right),  \\ \\
L'_0 = L_0 + \Delta L_0 = \frac{1}{3 \cdot 119} \left( 277 a - 283 b \right), \\ \\
L'_1 = L_1 + \Delta L_1 = \frac{1}{3 \cdot 119} \left( -185 a - 52 b \right), \\ \\
L'_2 = L_2 + \Delta L_2 = \frac{1}{3 \cdot 119} \left( -86 a + 26 b
\right).
\end{array}
\end{equation}

\noindent The $L_2 \neq 0$ particles decay via the operator ${\cal
O}_1$, changing the asymmetries in the proportions $\Delta L_0 =
\Delta L_2$ and $\Delta L_1 = -2 \Delta L_2$, as can be seen from
Eq. (5). Therefore, if $L_2 \longrightarrow 0$, one ends up with the
final values of the quantum numbers being

\begin{equation}
\begin{array}{l}
B_{0f} = \frac{2}{119} \left( a + 8 b \right),  \\ \\
L_{0f} = \frac{1}{119} \left( 121 a - 103 b \right), \\ \\
L_{1f} = -a.
\end{array}
\end{equation}

It should be noted that the final symmetry in $L_1$ is in the form
of the massive dark matter particles $(\Psi_6, \eta)$, and not in
the form of the massless fermions $\tilde{\Psi}$. (It is easily
shown in the following way that there is no asymmetry in
$\tilde{\Psi}$. One can assign an exactly conserved quantum number
$Z$ to $\tilde{\Psi}$ and $\tilde{\phi}_6$, with both these
particles having $Z=1$ and all other particles having $Z=0$. By
conservation of $Z$, and the fact that no asymmetry of $Z$ existed
initially, one has that $N_{\tilde{\Psi}} - N_{\tilde{\phi}_6} = 0$.
But eventually all the $\tilde{\psi}_6$ and their antiparticles
decay by $\tilde{\phi}^6 \longrightarrow \Psi_6 + \tilde{\Psi}$,
which drives  $N_{\tilde{\phi}_6}$ and thus $N_{\tilde{\Psi}}$ to
zero.)

Let us suppose that the primordial asymmetry generated during stage
1 is in the number of $\zeta$ particles. Since these have $X=2$ and
$B-L =1$, it follows that $a = 2b$. From this and Eq. (13), one has
that

\begin{equation}
\frac{L_{1f}}{B_{0f}} = -\frac{119}{10}.
\end{equation}

\noindent This is the present ratio of the number of dark matter
particles to the number of baryons in the universe. It is rather
remarkable feature of models of this type
\cite{bcf,adm-hdops,adm-sphaleron} that this ratio is predicted, and
therefore that the mass of the dark matter particle is predicted. Of
course, different grand unified models would give different
predictions.

Finally, we come to the question of the primordial asymmetry
generated during stage 1. One could imagine that this asymmetry was
in ordinary leptons, from the decay of superheavy right-handed
neutrinos $\Psi$, as has been much studied \cite{rhn-leptogenesis}.
However, that would give $a = 0$ and $b \neq 0$, which would yield
no dark matter asymmetry, according to Eq. (13). But asymmetries in
other species of particle can be generated in an analogous manner.
For example, suppose that the discrete $Z_N$ symmetry in Table I is
$Z_8$ and there exist complex scalar fields $S$ which are $SU(6)$
singlets and transform under $Z_8$ as $S \longrightarrow \omega^4 S
= - S$. Then, by Table I, one sees that the $S$ can have the Yukawa
coupling $\zeta \zeta S$ and also have explicit superheavy masses
from terms of the form $M_S^2 S^* S$ and $\Delta^2_S S \; S + H.c.$
The out-of-equilibrium decays of the superheavy $S$ particles can
generate a $\zeta-\overline{\zeta}$ asymmetry.

We conclude by observing that the model presented here is not
unique, but is meant to illustrate the general point that grand
unification, especially with large unitary groups, entails the
existence of Standard-Model-singlet fermions that could be the dark
matter. Such dark matter particles would be very difficult to
detect. For example, in the model presented here, the dark matter
particles have no Standard Model couplings except $O(v^2/M^{'2})$
couplings to the $Z$ boson through the mixing shown in Eq. (4). It
is characteristic of these scenarios that there will be extra $Z$
bosons at energies near the weak scale, and that the dark matter
will couple to these extra $Z$ bosons. Indeed, the primary means by
which the dark matter particles would be produced in accelerators
would be via the production of extra $Z$ bosons and their subsequent
decay into dark matter particle-antiparticle pairs.

\end{document}